\documentclass[aps,prd,preprint,tightenlines,superscriptaddress]{revtex4}
\usepackage{epsfig}
\usepackage{graphicx}
\usepackage{psfrag}
\usepackage{amsmath,amssymb}
\usepackage{colordvi}
\usepackage{color}
\usepackage{amsfonts}
\usepackage{enumerate}
\usepackage{slashed,bm}

\def \non {\nonumber}
\def \beq  {\begin{equation}}
\def \eeq  {\end{equation}}

\begin{document}

\title{On the renormalization of quasi parton distribution}
\author{Xiangdong Ji}
\affiliation{INPAC, Department of Physics and Astronomy, Shanghai Jiao Tong University, Shanghai, 200240, P. R. China}
\affiliation{Center for High-Energy Physics, Peking University, Beijing, 100080, P. R. China}
\affiliation{Maryland Center for Fundamental Physics, Department of Physics,  University of Maryland, College Park, Maryland 20742, USA}
\author{Jian-Hui Zhang}
\affiliation{INPAC, Department of Physics and Astronomy, Shanghai Jiao Tong University, Shanghai, 200240, P. R. China}
\affiliation{Institut f\"ur Theoretische Physik, Universit\"at Regensburg, \\
D-93040 Regensburg, Germany}
\vspace{0.5in}
\begin{abstract}

Recent developments showed that light-cone parton distributions can be studied by investigating the large momentum limit of the hadronic matrix elements of spacelike correlators, which are known as quasi parton distributions. Like a light-cone parton distribution, a quasi parton distribution also contains ultraviolet divergences and therefore needs renormalization. The renormalization of non-local operators in general is not well understood. However, in the case of quasi quark distribution, the bilinear quark operator with a straight-line gauge link appears to be multiplicatively renormalizable by the quark wave function renormalization in the axial gauge. We first show that the renormalization of the self energy correction to the quasi quark distribution is equivalent to that of the heavy-light quark vector current in heavy quark effective theory at one-loop order. Assuming this equivalence at two-loop order, we then show that the multiplicative renormalizability of the quasi quark distribution is true at two-loop order. 

\end{abstract}

\maketitle

\section{introduction}
Parton distribution functions (PDFs) characterize the structure of hadrons in terms of their fundamental constituents --- quark and gluon partons. They play a crucial role in describing high energy scattering experiments involving hadrons. According to the QCD factorization theorem~\cite{Mueller:1989hs,Sterman:1994ce,Collins:2011zzd}, the hadronic cross section for simple processes such as inclusive deep inelastic scattering and Drell-Yan processes can be computed as a convolution of the partonic cross section and the PDFs. Whereas the former is perturbatively calculable, the latter are intrinsically non-perturbative and difficult to compute. In the field-theoretic language, the PDFs are defined in terms of non-local light-cone correlators, 
which are intrinsically Minkowskian and therefore forbid a direct lattice simulation. Currently the extraction of PDFs relies on a suitable parametrization and fitting of the parameters to experimental or lattice data~\cite{Alekhin:2012ig,Gao:2013xoa,Radescu:2010zz,CooperSarkar:2011aa,Martin:2009iq,Ball:2012cx,negele}.

Recent developments~\cite{Ji:2013fga,Ji:2013dva,Xiong:2013bka,Lin:2014zya,Hatta:2013gta,Ma:2014jla,Ji:2014hxa,Ji:2014gla,Ji:2014lra} showed that the light-cone parton distribution can be directly extracted from the large momentum limit of the hadronic matrix element of a spacelike correlator, which is known as the quasi parton distribution. The quasi parton distribution does not have a real time dependence, and thus can be simulated on the lattice. Moreover, it has the same infrared (IR) behavior as the light-cone parton distribution. Its connection to the light-cone parton distribution can be established as a perturbative matching condition, which can also be viewed as a factorization. In Ref.~\cite{Xiong:2013bka}, we derived the matching condition at one-loop level for the non-singlet quark distribution. We showed explicitly that the soft divergences are canceled both for the quasi and for the light-cone distribution. The collinear divergences are the same for both distributions as well. We also presented the one-loop matching factor that transforms the frame dependence of the quasi distribution into the renormalization scale dependence of the light-cone distribution.

The factorization in Ref.~\cite{Xiong:2013bka} was given for the bare quasi quark distribution, where all fields and couplings entering the quasi distribution definition are bare ones. However, both the light-cone and the quasi distribution contain ultraviolet (UV) divergences and therefore need renormalization. In contrast to the renormalization of local operators, where all UV divergences can be removed by local counterterms, the renormalization of a non-local operator matrix element such as the parton distribution is rather distinct and less well understood. First, it contains UV divergences associated with coefficients that do not have a polynomial structure; second, there exist extra unphysical singularity structures, e.g. the light-cone singularities in the light-cone quark distribution or the axial singularities in the quasi quark distribution. These issues need to be properly addressed, in order to obtain a meaningful renormalized parton distribution. 

The renormalization properties of the light-cone quark distribution can be conveniently studied in the covariant Feynman gauge, as pointed out in Ref.~\cite{Collins:2011zzd}. Nevertheless, the choice of light-cone gauge provides a simpler picture in that the distribution reduces to the biproduct of two quark fields, and the renormalization also becomes more straightforward. The disadvantage of this gauge choice is that the treatment of light-cone singularities is not clear {\it a priori}. There have been a number of papers discussing the regularization of light-cone singularities, or of similar singularities in more general axial gauges~\cite{axial}. Given the similarity between the quasi and the light-cone distributions, it is natural that the renormalization of quasi quark distributions also becomes more straightforward in the axial gauge. There is, however, an important difference between the light-cone and the quasi distribution with respect to their UV behavior. In Ref.~\cite{Xiong:2013bka}, we have pointed out this difference: For the light-cone distribution, the momentum fraction is defined as $x=k^+/p^+$, where $k^+$ and $p^+$ are the plus-momenta for the quark in the loop and the initial quark, respectively. The light-cone momentum fraction is restricted to $[0,1]$, and the momentum $(k^+, k^-, k_\perp)$ can roughly be power-counted as $\sim(p^+, \Lambda^2/p^+, \Lambda)$ with $\Lambda$ being a UV cutoff. The UV divergences in the light-cone parton distribution are then genuine UV divergences to which the usual renormalization procedure can be applied.  In contrast, the momentum fraction for the quasi distribution is defined as $x=n\cdot k/n\cdot p$, where $n$ is a space-like vector, and is chosen to be along the $z$-direction in Ref.~\cite{Xiong:2013bka}. The self energy correction to the quasi distribution has the same UV behavior as in the light-cone distribution, since all components of the loop momentum are integrated over. However, the vertex correction behaves differently, since the quasi momentum fraction $x$, or equivalently, the $z$-component of the loop momentum $k^z$, is left unintegrated in the vertex correction. As discussed in Ref.~\cite{Xiong:2013bka}, the quasi momentum fraction is no longer restricted to $[0,1]$, it can extend from $[-\infty,+\infty]$. Therefore leaving $x$ unintegrated reduces the power of UV divergences, and leads to a UV convergent vertex correction at one-loop. Of course, the UV divergences appear when going beyond one-loop. However, as we will see, they can be subdivergences coming from subdiagrams only, not overall divergences. Consequently, all UV divergences in the (axial gauge) vertex correction can be removed by counterterms for subdiagrams from the interaction and therefore do not affect the renormalization of the quasi quark distribution; the renormalization of the quasi quark distribution then reduces to the renormalization of two quark fields in the axial gauge, provided that the non-trivial complication due to the choice of axial gauge does not matter. We explicitly show that this is indeed the case at two-loop order.

We will focus in particular on the unpolarized non-singlet quasi quark distribution. By investigating the one-loop corrections to the quasi quark distribution, we find an equivalence between the self energy correction of the quasi quark distribution and the correction to the heavy-light quark vector current in heavy quark effective theory (HQET), so that the UV divergences in the former can be renormalized as the renormalization of the latter. Based on this observation and the one-to-one correspondence between the two-loop diagrams, we then assume an equivalence between the UV divergences of the two-loop self energy in the quasi quark distribution and of the two-loop corrections of the heavy-light quark current, and discuss the two-loop UV divergence structure of the vertex correction and the renormalization of the quasi quark distribution. To simplify our calculation, we choose dimensional regularization for UV divergences throughout this paper. In this way, the linear divergence present in a cutoff regularization is ignored. The result presented in this paper therefore shall be viewed as a first step towards a full understanding of the renormalization property of the quasi distribution on the lattice.  

The rest of this paper is organized as follows. In Sec. 2, we briefly review the one-loop calculation performed in the axial gauge in Ref.~\cite{Xiong:2013bka}, and discuss its connection to the Feynman gauge computation. We also present a renormalization formula for the quasi quark distribution, and give the one-loop renormalization factor that renders the quasi quark distribution UV finite. In Sec. 3, we consider the two-loop correction and renormalization of the quasi quark distribution. Also some discussions on its renormalization beyond two-loop are given. Sec. 4 is our conclusion.

\section{one-loop correction and renormalization for the quasi quark distribution}
Let us start by recalling the definition of the quasi quark distribution. For the unpolarized quark density, it is given as~\cite{Ji:2013dva}
\beq\label{qUnpolDef}
   \tilde q(x, \mu, P^z) = \int^\infty_{-\infty} \frac{dz}{4\pi} e^{izk^z}  \langle P|\overline{\psi}(0, 0_\perp, z)
   \gamma^z \exp\left(-ig\int^{z}_0 dz' A^{z}(0, 0_\perp, z') \right)\psi(0) |P\rangle \ ,
\eeq
where we have chosen dimensional regularization for potential UV divergences, and $\mu$ denotes the renormalization scale dependence.

\begin{figure}[htbp]
\centering
\includegraphics[width=0.2\textwidth]{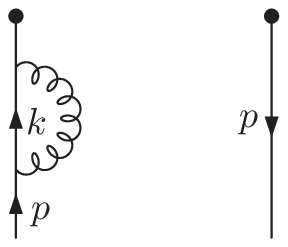}
\hspace{6em}
\includegraphics[width=0.2\textwidth]{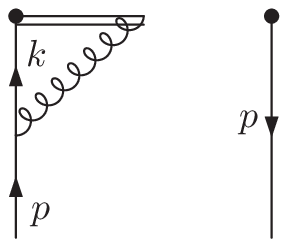}
\hspace{6em}
\includegraphics[width=0.2\textwidth]{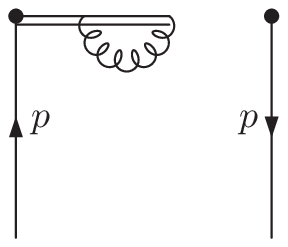}
\\
\vspace*{1.5em}
\hspace*{-.2em}
\includegraphics[height=0.165\textwidth]{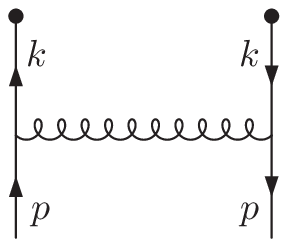}
\hspace*{6.2em}
\includegraphics[width=0.2\textwidth]{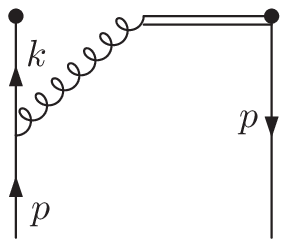}
\hspace{6em}
\includegraphics[width=0.2\textwidth]{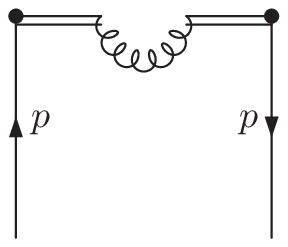}
\caption{One-loop diagrams for quasi quark distribution in Feynman gauge, the conjugate diagrams are not shown.}
\label{1loopFeyngauge}
\end{figure}

In Ref.~\cite{Xiong:2013bka}, we computed the one-loop correction to the unpolarized quasi quark distribution in the axial gauge $A^z=0$. The advantage of this gauge is that the gauge link in Eq.~(\ref{qUnpolDef}) reduces to unity, and the contributing Feynman diagrams become simple. The disadvantage is that it is non-covariant, and the prescription for the axial gauge singularity is not clear {\it a priori}. Since the definition Eq.~(\ref{qUnpolDef}) is gauge invariant, we can carry out the computation in any gauge. A convenient choice for the investigation of its renormalization properties is the covariant Feynman gauge, as mentioned in the Introduction. In the Feynman gauge, the contributing diagrams are given by those in Fig.~\ref{1loopFeyngauge}. Actually there is a straightforward correspondence between the diagrams in Fig.~\ref{1loopFeyngauge} and the individual terms in the axial gauge result. For example, the diagrams in the second row of Fig.~\ref{1loopFeyngauge} correspond to the contribution to the vertex correction from the $g^{\mu\nu}$ term, the $-(n^\mu (p-k)^\nu+n^\nu(p-k)^\mu)/n\cdot (p-k)$ term, and the $n^2(p-k)^\mu(p-k)^\nu/((n\cdot(p-k))^2$ term of the axial gauge gluon numerator, respectively. 

Let us write down the contribution of each Feynman gauge diagram. We start from the vertex corrections, {\it i.e.} the diagrams in the second row of Fig.~\ref{1loopFeyngauge}. The first diagram gives
\begin{align}
\Gamma_{11}&=\int\frac{d^4k}{(2\pi)^4}\bar u(p)(-ig t^a\gamma^\mu)\frac{i}{\slashed k-m}\gamma^z\frac{i}{\slashed k-m}(-igt^a\gamma_\mu)\frac{-i}{(p-k)^2}u(p)\delta\big(x-\frac{k^z}{p^z}\big)\non\\
&=-ig^2 C_F\int\frac{d^4k}{(2\pi)^4}\bar u(p) \Big[\frac{2\gamma^z}{(k^2-m^2)(p-k)^2} + \frac{8mk^z - 4k^z \slashed k}{(k^2-m^2)^2 (p-k)^2}\Big]u(p)\delta\big(x-\frac{k^z}{p^z}\big),
\end{align}
where the quark mass $m$ is introduced as in Ref.~\cite{Xiong:2013bka} to regularize the collinear divergences.

The above integrals can be computed in the same way as in Ref.~\cite{Xiong:2013bka}. After a Feynman parametrization and integration over $k^0$ and $\vec k_\perp$, we have the following contribution to the one-loop quasi distribution
\begin{align}
\tilde q_{11}&=\frac{\alpha_S C_F}{2\pi}\left\{ \begin{array} {ll} (x-1)\ln \frac{x-1}{x}+1\ , & x>1\ , \\ (1-x)\ln \frac{(p^z)^2}{m^2}+(1-x)\ln\frac{4x}{1-x}+1-\frac{2x}{1-x}\ , & 0<x<1\ , \\ (x-1)\ln \frac{x}{x-1}-1\ , & x<0\ . \end{array} \right.
\end{align}

The second diagram yields
\begin{align}\label{gamma12}
\Gamma_{12}&=\int\frac{d^4 k}{(2\pi)^4}\bar u(p) (igt^a)\frac{-i}{n\cdot(p-k)}\gamma^z\frac{i}{\slashed k-m}(-ig t^a\gamma^z)\frac{-i}{(p-k)^2}u(p)\delta\big(x-\frac{k^z}{p^z}\big)\non\\
&=-ig^2 C_F\int\frac{d^4 k}{(2\pi)^4}\bar u(p)\frac{\gamma^z(\slashed k+m)\gamma^z}{(p-k)^2(k^2-m^2)n\cdot(p-k)}u(p)\delta\big(x-\frac{k^z}{p^z}\big)\non\\
&=-ig^2 C_F\int\frac{d^4 k}{(2\pi)^4}\bar u(p)\frac{2k^z\gamma^z+\slashed k-m}{(p-k)^2(k^2-m^2)n\cdot(p-k)}u(p)\delta\big(x-\frac{k^z}{p^z}\big),
\end{align}
and leads to the one-loop correction (together with the conjugate diagram)
\begin{align}
\tilde q_{12}&=\frac{\alpha_S C_F}{2\pi}\left\{ \begin{array} {ll} -\frac{2x}{1-x}\ln \frac{x-1}{x}-\frac{1}{1-x}\ , & x>1\ , \\ \frac{2x}{1-x}\ln \frac{(p^z)^2}{m^2}+\frac{2x}{1-x}\ln\frac{4x}{1-x}+1-\frac{x}{1-x}\ , & 0<x<1\ , \\ -\frac{2x}{1-x}\ln \frac{x}{x-1}+\frac{1}{1-x}\ , & x<0\ . \end{array} \right.
\end{align}

Note that there is no UV divergence in the above results. This is because the momentum fraction $x$ is left unintegrated. As $x$ can extend between $[-\infty,+\infty]$, leaving it unintegrated reduces the power of UV divergence, and therefore leads to a UV convergent result in the one-loop vertex correction. If one integrates over the momentum fraction $x$, as is done for the self energy diagrams in the first row of Fig.~\ref{1loopFeyngauge}, one will find a logarithmic UV divergence, in accordance with the usual UV power counting.

The third diagram gives (note $n^2=-1$)
\begin{align}\label{gamma13}
\Gamma_{13}&=\int\frac{d^4k}{(2\pi)^4}\bar u(p)\gamma^z (ig t^a)(-ig t^a)n^2 \frac{-i}{n\cdot(p-k)}\frac{i}{n\cdot(p-k)}\frac{-i}{(p-k)^2} u(p)\delta\big(x-\frac{k^z}{p^z}\big)\non\\
&=ig^2 C_F\int\frac{d^4k}{(2\pi)^4}\bar u(p)\frac{\gamma^z}{(p-k)^2[n\cdot(p-k)]^2}u(p)\delta\big(x-\frac{k^z}{p^z}\big),
\end{align}
which leads to the following result
\begin{align}
\tilde q_{13}&=\frac{\alpha_S C_F}{2\pi}\left\{ \begin{array} {ll} \frac{1}{1-x}\ , & x>1\ , \\ -\frac{1}{1-x}\ , & 0<x<1\ , \\ -\frac{1}{1-x}\ , & x<0\  \end{array} \right.
\end{align}
in dimensional regularization.

Now let us look at the self energy diagrams in the first row of Fig.~\ref{1loopFeyngauge}, where we integrate over all components of the loop momentum. Since we are interested in the renormalization properties of the quasi distribution, we need the UV divergent part of the diagrams only. The first diagram is the usual Feynman gauge quark self energy diagram, it leads to the following contribution to the wave function renormalization factor $\tilde Z_F^{(1)}$ in dimensional regularization ($D=4-2\epsilon$) and $\overline{MS}$ scheme
\beq\label{z11result}
\tilde Z_{11}=-\frac{\alpha_S C_F S_\epsilon}{4\pi}\frac{1}{\epsilon},
\eeq
where $S_\epsilon=(4\pi)^\epsilon/\Gamma(1-\epsilon)$ accounts for the constant associated with $1/\epsilon$ in the $\overline{MS}$ scheme.
The second diagram yields 
\begin{align}\label{z12}
\Sigma_{12}&=\int\frac{d^4 k}{(2\pi)^4}(-igt^a)\frac{-i}{n\cdot(p-k)}\frac{i}{\slashed k-m}(-ig t^a\gamma^z)\frac{-i}{(p-k)^2},
\end{align}
In dimensional regularization, this leads to the following contribution to $\tilde Z_F^{(1)}$
\beq\label{z12result}
\tilde Z_{12}=\frac{\alpha_S C_F S_\epsilon}{2\pi}\frac{1}{\epsilon}.
\eeq
To understand how the Wilson line self energy diagram contributes, let us expand the quasi distribution definition in coordinate space Eq.~(\ref{qUnpolDef}) to $\mathcal O(g^2)$. For convenience, let us separate the gauge link as
\beq\label{gaugelinksep}
\mathcal P e^{-ig\int_0^z dz' n\cdot A(0,0_\perp, z')}=[\mathcal P e^{-ig\int_z^\infty dz' n\cdot A(0,0_\perp, z')}]^\dagger \mathcal P e^{-ig\int_0^\infty dz' n\cdot A(0,0_\perp, z')},
\eeq
and associate the first factor on the r.h.s. to $\bar\psi(z)$ and the second to $\psi(0)$ in order to form gauge invariant quark fields. It is easy to see that the Wilson line self energy diagram arises when the $\mathcal O(g^2)$ term comes from one of the path-ordered exponentials on the r.h.s. of Eq.~(\ref{gaugelinksep}) only. If each of the exponentials contributes an $\mathcal O(g)$ term, this leads to the last diagram in Fig.~\ref{1loopFeyngauge}. Suppose we expand the second exponential to $\mathcal O(g^2)$, we have  
\begin{align}\label{z13}
&\int\frac{dz}{4\pi}e^{ix p^z z}\langle p|\bar\psi(0,0_\perp,z)\gamma^z \frac{-g^2 C_F}{2}\int_0^\infty dz' dz'' \mathcal P (n\cdot A(0,0_\perp, z')n\cdot A(0, 0_\perp, z''))\psi(0)|p\rangle\non\\
&=\int\frac{dz}{4\pi}e^{ix p^z z}\langle p|\bar\psi(0,0_\perp,z)\gamma^z \frac{-g^2 C_F}{2}\int_0^\infty dz'  dz'' \int\frac{d^4 k}{(2\pi)^4}\frac{-i n^2}{(p-k)^2}e^{i(p^z-k^z)(z'-z'')}\psi(0)|p\rangle\non\\
&=-\frac{ig^2 C_F}{2}\int\frac{d^4 k}{(2\pi)^4}\frac{1}{(p-k)^2[n\cdot(p-k)]^2}\delta(1-x).
\end{align}
The expansion of the first exponential yields the same contribution. This derivation also gives a natural prescription for the double pole at $n\cdot(p-k)=0$. From the $z', z''$ integration above, one can see that to achieve well-defined Wilson line propagators in momentum space, one must have a $+i\varepsilon$ prescription for one $1/n\cdot(p-k)$ propagator, and a $-i\varepsilon$ for the other. This leads to the prescription $1/\big((n\cdot(p-k))^2+\varepsilon^2\big)$ for the double pole $1/(n\cdot(p-k))^2$. The above derivation is also formally consistent with the intuitive understanding of the third diagram in the first row of Fig.~\ref{1loopFeyngauge}, where the Wilson line propagator $1/n\cdot (p-k)$ and the corresponding self energy correction are similar to those for a quark matter field in the on-shell and zero momentum limit, its contribution can therefore be extracted analogously by taking the residue of the self energy correction in the zero momentum limit, which is given by
\begin{align}\label{sigma13}
&\lim_{l\to 0}n^\mu \frac{\partial}{\partial l^\mu}\int\frac{d^4 k}{(2\pi)^4}(-i)(-ig t^a)(-ig t^a)n^2\frac{-1}{(p-k)^2 n\cdot(p-k-l)}\non\\
&=-ig^2 C_F\int\frac{d^4 k}{(2\pi)^4}\frac{1}{(p-k)^2 [n\cdot(p-k)]^2}.
\end{align}
From Eq.~(\ref{z13}) or Eq.~(\ref{sigma13}) and the prescription for the double pole, we obtain the following contribution to $\tilde Z_F^{(1)}$ from the Wilson line self energy diagram
\beq\label{z13result}
\tilde Z_{13}=\frac{\alpha_S C_F S_\epsilon}{2\pi}\frac{1}{\epsilon}.
\eeq


Now we can write down the renormalization of quasi quark distribution. The quasi distribution can be renormalized as follows
\beq\label{pdfrenorm}
\tilde q_R(x,p^z)=\int \frac{dy}{|y|} Z(\frac{x}{y})\tilde q(y, p^z),
\eeq
where the subscript $R$ denotes the renormalized quasi distribution, $Z(x/y)$ is the renormalization factor required to cancel the UV divergence. The above relation can be rewritten as
\beq\label{pdfrenorm1}
\tilde q_R(x,p^z)=\int \frac{dy}{|y|}[\tilde Z_F Z(\frac{x}{y})][\tilde Z_F^{-1}\tilde q(y, p^z)],
\eeq  
where we separate the field wave function renormalization constant $\tilde Z_F$ such that the factor in the second bracket is the quasi distribution with renormalized fields and can be computed using the standard Feynman rules, with the counterterms from the Lagrangian also being taken into account. The factor in the first bracket then cancels the remained UV divergence in the parton distribution. 

From the one-loop results above, we can see that there is no UV divergence in the vertex correction (the second row of Fig.~\ref{1loopFeyngauge}). The UV divergence arises only from the diagrams in the first row of Fig.~\ref{1loopFeyngauge}, the sum of which corresponds to the quark self energy in the axial gauge. Therefore, at one-loop level one needs to renormalize the self energy only, and the renormalization constant is given by
\begin{align}\label{1looppdfrenconst}
Z(\eta)=\delta(\eta-1)(1+Z^{(1)})=\delta(\eta-1)-\frac{3\alpha_S C_F S_\epsilon}{4\pi}\frac{1}{\epsilon}\delta(\eta-1).
\end{align}
An interesting observation is that the diagrams in the first row of Fig.~\ref{1loopFeyngauge} are in one-to-one correspondence with the one-loop correction to the heavy-light quark vector current $q\gamma^z Q$ in the HQET. Also the Wilson line propagator $1/n\cdot k$ is the same as $1/v\cdot k$ in the heavy quark propagator, except that the vector $n$ in the former is space-like, whereas the velocity $v$ in the latter is time-like. However, as we see from Eqs.~(\ref{z11result}), (\ref{z12}), (\ref{z12result}), (\ref{z13}), (\ref{sigma13}), (\ref{z13result}), the UV divergences are actually identical for the one-loop diagrams in both cases (up to a factor of 2 since the contributing diagrams for quasi quark distribution are twice of those for heavy-light quark vector current). Therefore, the one-loop renormalization for quasi quark distribution requires the renormalization of self energy diagrams only, which is equivalent to the renormalization of the heavy-light quark vector current in HQET.

\begin{figure}[htbp]
\begin{minipage}[htbp]{\textwidth}
\centering
\includegraphics[width=0.195\textwidth]{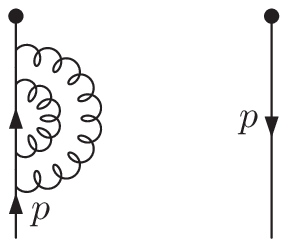}
\hspace{4.9em}
\includegraphics[width=0.23\textwidth]{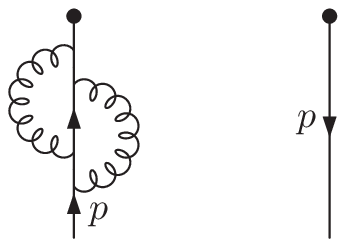}
\hspace{4.9em}
\includegraphics[width=0.195\textwidth]{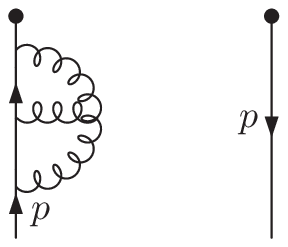}
\\
\vspace{1.5em}
\hspace{-.3em}
\includegraphics[height=0.165\textwidth]{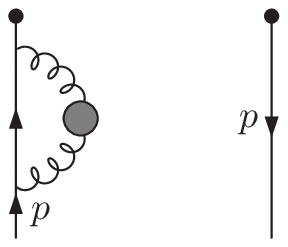}
\hspace{6.5em}
\includegraphics[width=0.195\textwidth]{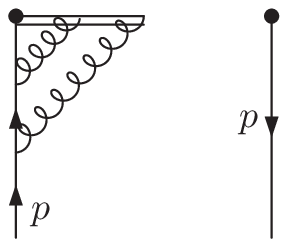}
\hspace{4.9em}
\includegraphics[width=0.195\textwidth]{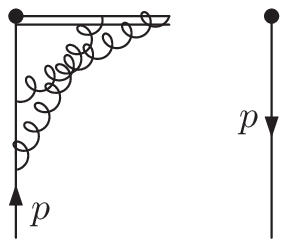}
\\
\vspace{1.5em}
\hspace{-1.8em}
\includegraphics[height=0.165\textwidth]{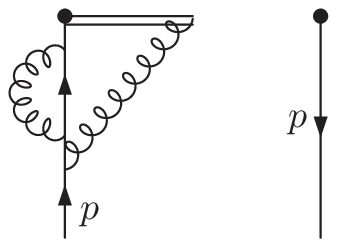}
\hspace{6.5em}
\includegraphics[width=0.195\textwidth]{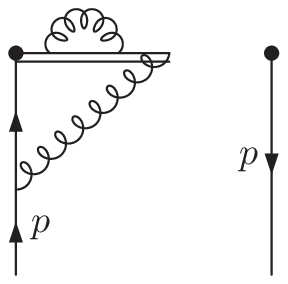}
\hspace{4.9em}
\includegraphics[width=0.195\textwidth]{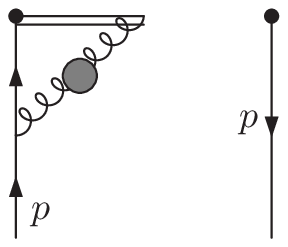}
\\
\vspace{1.5em}
\hspace{-2.2em}
\includegraphics[height=0.165\textwidth]{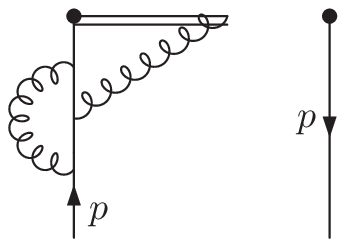}
\hspace{6.5em}
\includegraphics[width=0.195\textwidth]{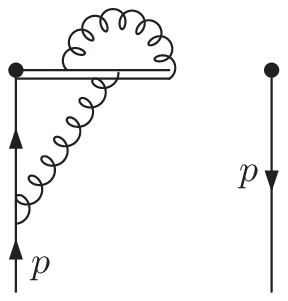}
\hspace{4.9em}
\includegraphics[width=0.195\textwidth]{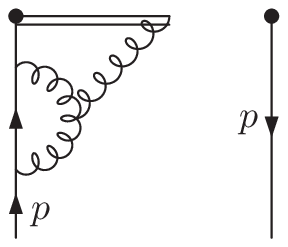}
\\
\vspace{1.5em}
\hspace{1.2em}
\includegraphics[height=0.165\textwidth]{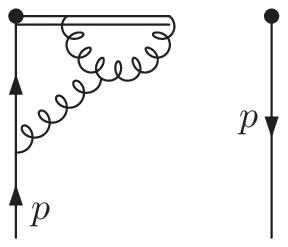}
\end{minipage}
\caption{Two-loop self energy diagrams for quasi quark distribution in Feynman gauge. The disjoint and conjugate diagrams are not shown. The two-loop Wilson line self energy diagrams are the same as the first four diagrams for quark self energy, but with gluons attached to the Wilson line.}
\label{2loopFeyngaugeSE}
\end{figure}

\section{Renormalization at two-loop}
In this section we consider the two-loop correction to the quasi quark distribution. Following our discussion of one-loop diagrams, we can divide the two-loop Feynman gauge diagrams into two classes: one corresponds, as the first row of Fig.~\ref{1loopFeyngauge}, to the two-loop diagrams for the heavy-light quark current in HQET~\cite{Ji:1991pr}, or to the two-loop quark self energy in the axial gauge; the other corresponds, as the second row of Fig.~\ref{1loopFeyngauge}, to the two-loop vertex correction. The first class diagrams are depicted in Fig.~\ref{2loopFeyngaugeSE}, where we do not show the disjoint diagrams with two one-loop diagrams separated from each other. Also conjugate diagrams are not shown. The Wilson line self energy diagrams are the same as the first four quark self energy diagrams, but with gluons attached to the Wilson line.

\begin{figure}[htbp]
\centering
\hspace{0em}
\includegraphics[width=0.195\textwidth]{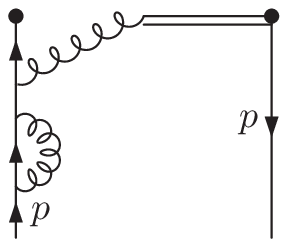}
\hspace{6.1em}
\includegraphics[width=0.2\textwidth]{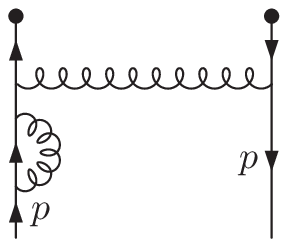}
\hspace{4.2em}
\includegraphics[width=0.245\textwidth]{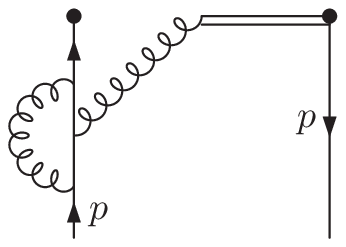}
\\
\vspace*{1.5em}
\hspace*{-2em}
\includegraphics[height=0.165\textwidth]{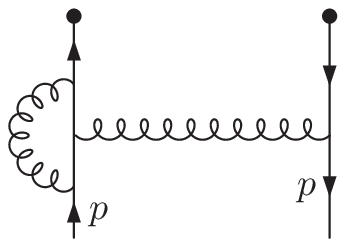}
\hspace*{4.3em}
\includegraphics[width=0.245\textwidth]{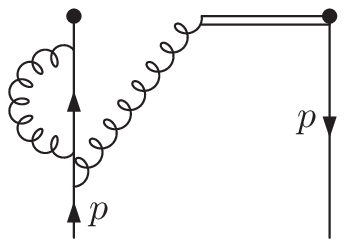}
\hspace{4.1em}
\includegraphics[width=0.245\textwidth]{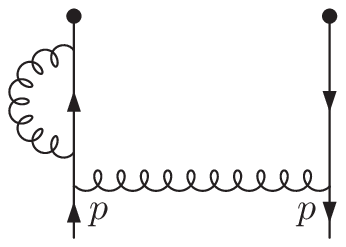}
\\
\vspace*{1.5em}
\hspace*{-.3em}
\includegraphics[height=0.165\textwidth]{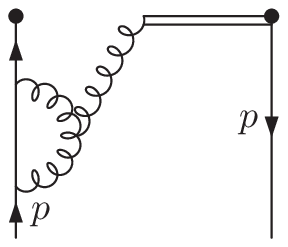}
\hspace*{6.2em}
\includegraphics[width=0.2\textwidth]{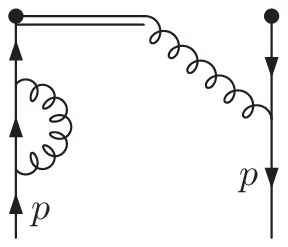}
\hspace{6em}
\includegraphics[width=0.2\textwidth]{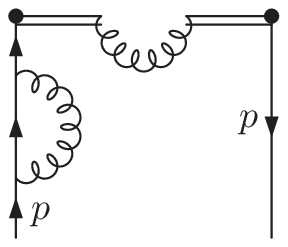}
\\
\vspace*{1.5em}
\hspace*{-.3em}
\includegraphics[height=0.165\textwidth]{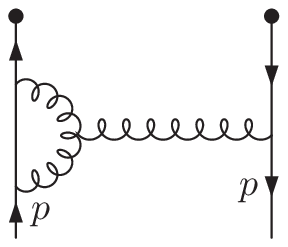}
\hspace*{6.2em}
\includegraphics[width=0.2\textwidth]{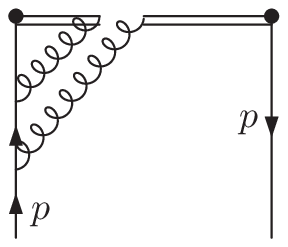}
\hspace{6em}
\includegraphics[width=0.2\textwidth]{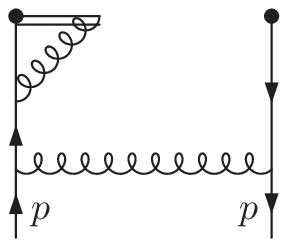}
\\
\vspace*{1.5em}
\hspace*{-.3em}
\includegraphics[height=0.165\textwidth]{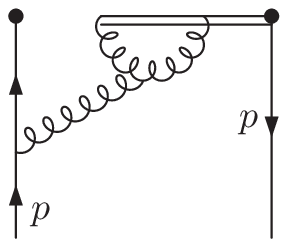}
\hspace*{6.2em}
\includegraphics[width=0.2\textwidth]{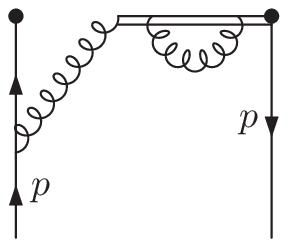}
\hspace{6em}
\includegraphics[width=0.2\textwidth]{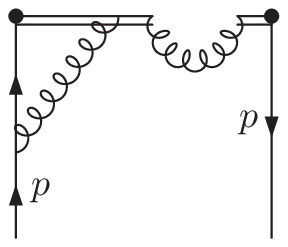}
\caption{Feynman gauge two-loop vertex diagrams that contain subdivergences, the conjugate diagrams are not shown (to be continued).}
\label{2loopFeyngauge1}
\end{figure}
\begin{figure}[htbp]
\centering
\hspace{-.3em}
\includegraphics[height=0.205\textwidth]{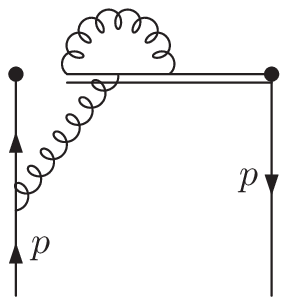}
\hspace*{6.1em}
\includegraphics[width=0.2\textwidth]{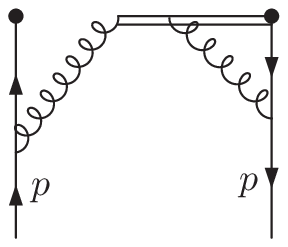}
\hspace{6em}
\includegraphics[width=0.2\textwidth]{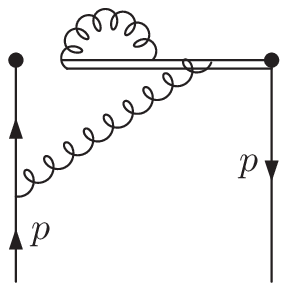}
\\
\vspace*{1.5em}
\hspace*{-.3em}
\includegraphics[height=0.185\textwidth]{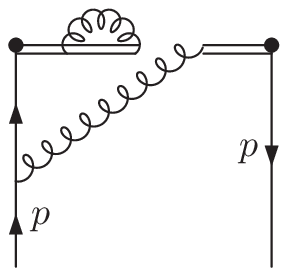}
\hspace*{6.2em}
\includegraphics[width=0.2\textwidth]{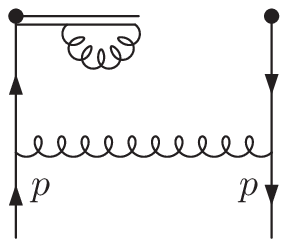}
\hspace{6em}
\includegraphics[width=0.2\textwidth]{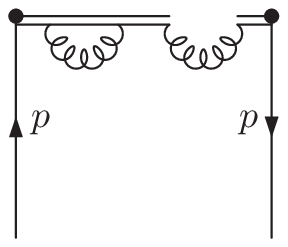}
\\
\vspace*{1.5em}
\hspace*{-.3em}
\includegraphics[height=0.2\textwidth]{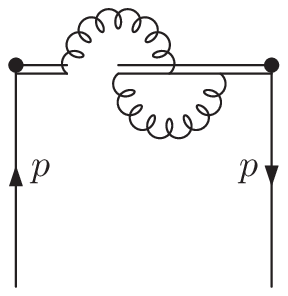}
\hspace*{6.2em}
\includegraphics[width=0.2\textwidth]{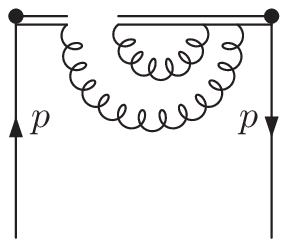}
\hspace{6em}
\includegraphics[width=0.2\textwidth]{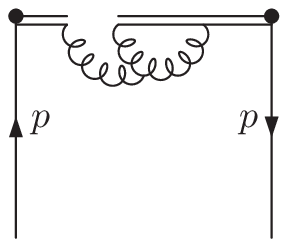}
\\
\vspace*{1.5em}
\hspace*{-.3em}
\includegraphics[height=0.165\textwidth]{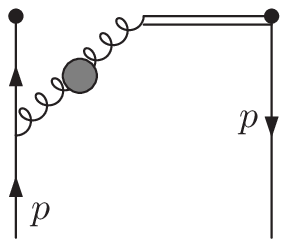}
\hspace*{6.2em}
\includegraphics[width=0.2\textwidth]{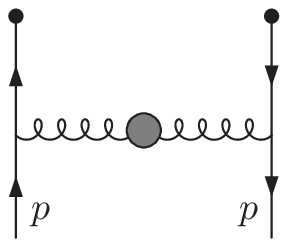}
\hspace{6em}
\includegraphics[width=0.2\textwidth]{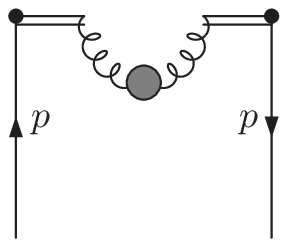}
\caption{Feynman gauge two-loop vertex diagrams that contain subdivergences (continued).}
\label{2loopFeyngauge2}
\end{figure}

As in the one-loop case, the two-loop diagrams in Fig.~\ref{2loopFeyngaugeSE} are in one-to-one correspondence with the two-loop diagrams for the heavy-light quark current in HQET, we can therefore renormalize the UV divergences in Fig.~\ref{2loopFeyngaugeSE} in the same way as is done for the two-loop renormalization of heavy-light quark current~\cite{Gimenez:1991bf}. Based on the equivalence of their one-loop corrections, in the following we will assume they have the same UV divergences also at two-loop level, and focus on the two-loop renormalization property of the quasi quark distribution. The UV divergent two-loop contribution to the heavy-light quark current has been analyzed in Refs.~\cite{Gimenez:1991bf,Ji:1991pr}.

We now consider the two-loop vertex diagrams. At two-loop, the overall UV divergence of a vertex diagram comes from the region where both loop momenta, denoted as $k_1$ and $k_2$, are large. When either $k_1$ or $k_2$ is large while the other is small, a subdivergence arises. According to our definition of quasi distribution, the momentum fraction $x=k_1^z/p^z$, or equivalently $k_1^z$, is left unintegrated in the two-loop vertex correction. Since $k_1^z$ can go to $\pm\infty$, leaving it unintegrated reduces the power of UV divergence, a simple power counting tells us that no overall UV divergence appears in the two-loop vertex diagrams. However, there exist subdivergences. In Figs.~\ref{2loopFeyngauge1} and~\ref{2loopFeyngauge2} we show all two-loop vertex diagrams that contain subdivergences. We are interested in the renormalization property of the diagrams, and therefore will focus on their UV behavior. Given the one-loop results in the previous section, the subdivergences in these diagrams can be computed straightforwardly. In Table~\ref{2loopvertUV} we list the UV divergent parts of all the two-loop vertex diagrams.

Summing over all the contributions in Table~\ref{2loopvertUV}, we have
\beq\label{gamma2}
\Gamma^{(2)}=\frac{\alpha_S}{12\pi\epsilon}(11C_A+9C_F-2n_f)\tilde q^{(1)},
\eeq
where we have used $T_f=1/2$ for the $SU(N)$ gauge group, $C_F=(N^2-1)/2N$, $C_A=N$, $n_f$ is the number of active quark flavors, and $\tilde q^{(1)}=\tilde q_{11}+\tilde q_{12}+\tilde q_{13}$. Although each contribution in Table~\ref{2loopvertUV} has a distinct divergence structure, the sum of them is proportional to the one-loop vertex correction $\tilde q^{(1)}$. This indicates that all UV divergences in the axial gauge vertex correction can be removed by counterterms from the interaction~(for the UV counterterms in the axial gauge see e.g. Ref.~\cite{axial}) and therefore do not affect the renormalization of the quasi quark distribution. The renormalization of the quasi quark distribution is then equivalent to the renormalization of the two separate quark fields in the axial gauge. Expanding the renormalization formula Eq.~(\ref{pdfrenorm1}) to two-loop level, we can write the two-loop renormalization factor for the quasi quark distribution as
\beq\label{2looprenfac}
Z^{(2)}(\eta)=\big[\big(\frac{\alpha_S}{4\pi}\big)^2 S_\epsilon^2\big(\frac{a}{\epsilon^2}+\frac{b}{\epsilon}\big)+(Z^{(1)})^2\big]\delta(\eta-1\big),
\eeq
where $a, b$ come from the renormalization of two-loop self energy diagrams in Fig.~\ref{2loopFeyngaugeSE}. As is well-known, the double pole term $a/\epsilon^2$ can be derived from the single pole at one-loop. Suppose we write a renormalization factor $Z$ as a series expansion
\beq
Z=\sum_{i=1}^\infty \sum_{j=1}^i {\big(\frac{\alpha_S}{4\pi}\big)}^i \frac{1}{\epsilon^j} Z_{i,j}.
\eeq
For the one-loop self energy diagrams in Fig.~\ref{1loopFeyngauge}, the renormalization factor in a general covariant gauge is given as~\cite{Groote:1996xb} ($\xi$ is a gauge parameter, and the Feynman gauge is obtained for $\xi=1$)
\beq
Z_{1,1}=2Z_{1,1}^{\rm {HQET}}=(-\xi C_F+(3-\xi)C_F)+2\xi C_F=3C_F,
\eeq
which is twice the renormalization factor of heavy-light quark current in HQET, since the contributing diagrams to $Z_{1,1}$ are twice those to the heavy-light quark current.

\begin{table}
\renewcommand{\arraystretch}{1.3}
\centering
\begin{tabular}{|c|c|c|c|}
\hline
Fig.~\ref{2loopFeyngauge1} & UV divergence & Fig.~\ref{2loopFeyngauge2} & UV divergence \\
\hline
1, 2, 8, 9 (+conj.) & $-\frac{\alpha_SC_F}{4\pi\epsilon}\tilde q^{(1)}$ & 1 (+conj.) & $-\frac{\alpha_S}{2\pi\epsilon}(C_F-\frac{1}{2}C_A)\tilde q_{12}$ \\
3, 4, 7, 10 (+conj.) & $\frac{\alpha_S}{4\pi\epsilon}(C_F+C_A)(2\tilde q_{11}+\tilde q_{12})$ & 2 (+conj.) & $\frac{\alpha_SC_F}{4\pi\epsilon}\tilde q_{12}$ \\
5, 6 (+conj.) & $-\frac{\alpha_SC_F}{4\pi\epsilon}(2\tilde q_{11}+\tilde q_{12})$ & 3, 4, 5 (+conj.) & $\frac{\alpha_SC_F}{2\pi\epsilon}(\tilde q_{11}+\tilde q_{12})$ \\
11, 12 (+conj.) & $\frac{\alpha_SC_F}{4\pi\epsilon}(2\tilde q_{11}+\tilde q_{12})$ & 6 (+conj.) & $\frac{\alpha_SC_F}{\pi\epsilon}\tilde q_{13}$ \\
13 (+conj.) & 0 & 7 (+conj.)  & $-\frac{\alpha_S}{2\pi\epsilon}(2C_F-C_A)\tilde q_{13}$ \\
14 (+conj.) & $\frac{\alpha_SC_F}{2\pi\epsilon}\tilde q_{12}$ & 8 (+conj.) & $\frac{\alpha_SC_F}{2\pi\epsilon}\tilde q_{13}$ \\
15 (+conj.) & $\frac{\alpha_SC_F}{2\pi\epsilon}\tilde q_{13}$ & 9 (+conj.) & 0 \\
 & & 10 (+conj.), 11, 12 & $\frac{\alpha_S}{4\pi\epsilon}(\frac{5}{3}C_A-\frac{4}{3}T_f n_f)\tilde q^{(1)}$ \\
\hline
\end{tabular}
\caption{Results for the UV divergent parts of two-loop vertex diagrams of quasi quark distribution. (+conj.) means including the contribution of conjugate diagrams.}
\label{2loopvertUV}
\end{table}

The renormalization group equation then leads to the following relation~\cite{Pascual:1984zb,Gimenez:1991bf,Groote:1996xb}
\beq
Z_{2,2}=-\frac{1}{4}(-2Z_{1,1}-\frac{22}{3}C_A+\frac{4}{3}n_f)Z_{1,1}=\frac{9}{2}C_F^2+\frac{11}{2}C_A C_F-n_f C_F,
\eeq
where we have used the fact that at one-loop $Z_{1,1}$ is independent of the gauge parameter $\xi$. The coefficient $a$ in Eq.~(\ref{2looprenfac}) is therefore given by
\beq
a=-\big(\frac{9}{2}C_F^2+\frac{11}{2}C_A C_F-n_f C_F\big).
\eeq
The single pole coefficient $b$ can be read off from the two-loop result for the heavy-light quark current in HQET in Refs.~\cite{Politzer:1988wp,Gimenez:1991bf,Ji:1991pr}
\beq
b=-\frac{127}{9}-\frac{28}{27}\pi^2+\frac{10}{9}n_f.
\eeq
Beyond two-loop, it is still true that the vertex contribution contains no overall divergence, but subdivergences only, as can be seen from the power counting above. In the axial gauge where the gauge link becomes unity, all these subdivergences can be removed by counterterms from the interaction, provided that the non-trivial complication due to the choice of axial gauge does not matter. The renormalization of the quasi quark distribution then reduces to the renormalization of two separate quark fields in the axial gauge. This remains, however, to be checked by explicit computations beyond two-loop order.  

In the above discussion, we choose dimensional regularization for UV divergences. It is interesting to investigate the renormalization property of the quasi distribution in the presence of a cutoff regulator used in Ref.~\cite{Xiong:2013bka} to mimic lattice setting. At one-loop, the cutoff regulator, denoted as $\Lambda$, introduces a linear divergence in $\tilde q_{13}$ and $\tilde Z_{13}$~\cite{Xiong:2013bka}. As discussed in Sec. 2, the prescription for the axial gauge double pole at $x=1$ is given by $1/((1-x)^2+\varepsilon^2)$. This leads to the following result for $\tilde Z_{13}$
\beq
\tilde Z_{13}=\frac{\alpha_S C_F}{2\pi}(\ln\Lambda^2-\frac{\pi}{\varepsilon}\frac{\Lambda}{p^z}).
\eeq
Since the vertex correction also contains a linear divergence, the one-loop renormalization factor for the quasi distribution now becomes
\beq
Z(\eta)=\delta(\eta-1)(1-\frac{\alpha_S C_F}{4\pi}(3\ln\Lambda^2-\frac{2\pi}{\varepsilon}\frac{\Lambda}{p^z}))-\frac{\alpha_S C_F}{2\pi}\frac{1}{(1-\eta)^2+\varepsilon^2}\frac{\Lambda}{p^z},
\eeq
where the $\varepsilon$-dependence drops out when integrating over a smooth function of $\eta$, as discussed in Ref.~\cite{Xiong:2013bka}.  We do not discuss the renormalization with a cutoff regulator beyond one-loop, since the use of a cutoff regulator is ambiguous at higher-loop order. Moreover, there is no unambiguous way to implement a cutoff gauge invariantly in the continuum, although this can be achieved on a discretized lattice.

\section{Conclusion}
We have discussed the renormalization property of the quasi parton distribution, focusing in particular on the unpolarized quasi quark distribution in the non-singlet case. In this case, the bilinear quark operator with a straight-line gauge link appears to be multiplicatively renormalizable by the quark wave function renormalization constant in the axial gauge. We explicitly showed that this is true at two-loop order. In the covariant Feynman gauge the self energy diagrams are in one-to-one correspondence with the loop diagrams for a heavy-light quark vector current in HQET at the same order, thus the renormalization of self energy correction can be carried out as  renormalization of the heavy-light quark vector current in HQET. The vertex correction of quasi quark distribution contains subdivergences only, which can be removed by counterterm for subdiagrams. Such features are expected to hold also beyond two-loop order, but remain to be checked by explicit computations.

We thank A. Sch\"afer for helpful discussions and comments on the manuscript. This work was partially supported by the U.S. Department of Energy via grants DE-FG02-93ER-40762, a grant (No. 11DZ2260700) from the Office of
Science and Technology in Shanghai Municipal Government, grants from National Science Foundation of China (No. 11175114, No. 11405104), and a DFG grant SCHA 458/20-1.


\end{document}